\documentclass[epsfig, usenatbib]{mn2e}
\usepackage{graphicx}
\usepackage{amssymb}

\title{Inter--cluster filaments in a $\Lambda$CDM Universe}
\author[J.\ M.\ Colberg {\it et al.}]
       {J\"org M.\ Colberg,$^{1, 2}$ K.\ Simon Krughoff,$^1$ Andrew J.\ Connolly$^1$\\ 
        $^1$ University of Pittsburgh, 3941 O'Hara Street, 
             100 Allen Hall, Pittsburgh PA 15260, USA\\
        $^2$ astro@jmcolberg.com}
\date{Accepted 200? ???? ??.
      Received 200? ???? ??;
      in original form 200?  xx}

\pagerange{\pageref{firstpage}--\pageref{lastpage}}
\pubyear{2005}

\begin{document}
\bibliographystyle{mn2e}

\maketitle

\label{firstpage}

\begin{abstract}
The large--scale structure (LSS) in the Universe comprises a complicated 
filamentary network of matter. We study this network using a high--resolution 
simulation of structure formation in a $\Lambda$ Cold Dark Matter cosmology. 
We investigate the distribution of matter between neighbouring 
large haloes whose masses are comparable to massive clusters of galaxies. We 
identify a total of 228 filaments between neighbouring clusters. Roughly half 
of the filaments are either warped or lie off the cluster--cluster axis. We 
find that straight filaments, on the average, are shorter than warped ones. 
Close cluster pairs with separation of 5\,$h^{-1}$\,Mpc or less are always 
connected by a filament. At separations between 15\,$h^{-1}$\,Mpc and 
20\,$h^{-1}$\,Mpc about a third of cluster pairs are connected by a filament. 
On average, more massive clusters are connected to more filaments than less 
massive ones. This finding indicates that the most massive clusters form at the 
intersections of the filamentary backbone of LSS. For straight filaments, we 
compute mass profiles. Radial profiles show a fairly well--defined radius, $r_s$, 
beyond which the profiles follow an $r^{-2}$ power law fairly closely. For 
the majority of filaments, $r_s$ lies between 1.5\,$h^{-1}$\,Mpc and 
2.0\,$h^{-1}$\,Mpc. The enclosed overdensity inside $r_s$ varies from a 
few times up to 25 times the mean density, independent of the length of the filament.
Along the filaments' axes, material is not distributed uniformly. Towards 
the clusters, the density rises, indicating the presence of the cluster infall 
regions. Filaments have been suggested to cause possible alignments between 
neighbouring clusters. Looking at the nearest neighbour for each cluster, we 
find that, up to a separation of about 15\,$h^{-1}$\,Mpc, there is a filament 
present that could account for alignment. In addition, we also find some 
sheet--like connections between clusters. In roughly a fifth of all 
cluster--cluster connections where we could not identify a filament or sheet, 
projection effects lead to filamentary structures in the projected mass distribution. 
\end{abstract}


\section{Introduction}

Large galaxy redshift surveys such as the Sloan Digital Sky Survey (York et al. 2000)
and the 2dFGRS (Colless et al. 2001) and N--body simulations of 
cosmic structure formation (for example \citet{jenkins98}, \citet{wambsganss04}) show a complicated 
network of matter. In redshift surveys, the galaxies line up preferentially around 
roundish, almost empty regions -- so--called voids. Clusters of galaxies have very 
prominent positions in the network: They lie at the intersections of filaments. In 
N--body simulations, this trend is even more pronounced. The network of haloes is 
clearly followed by the more diffusely distributed component of the dark matter 
that, at $z=0$, has not collapsed into haloes.

Describing this network is no easy task. There are a wide variety of 
statistical and topological tools to compare observations and theoretical models. 
Amongst these are the two--point (and higher--order) correlation function(s) 
\citep[e.g.][]{peebles80,peebles75}, minimal spanning trees 
\cite[see e.g.][]{barrow85,krzewina96,bhavsar96}, the genus statistics \citep{gott86}, 
Minkowski functionals \citep[that include the genus;][]{mecke94}, shape statistics 
\citep[see e.g.][]{babul92,luo95,luo96},
and shapefinder statistics derived from Minkowski functionals 
\citep[][]{sahni98,sheth03,shandarin04}. 
These tools have all been applied to galaxy catalogues and N--body
simulations and have been very useful in comparing observations and theory.

These tools have been somewhat less helpful in describing the pattern of
large--scale structure as far as filaments and sheets/walls are concerned. 
Minkowski functionals and especially their shapefinder cousins have been used
to try to measure how thick or long filaments are. By construction, for simple 
toy models like ideal filaments (cylinders) or sheets (planes), shape statistics 
give very simple answers. Because of the complexity of the network of matter in
surveys or simulations the answers for these cases are usually more 
difficult to interpret than in the toy model simulations. 
Recently, \citet{bharadwaj04} investigated 
filaments in the Las Campanas Redshift Survey using shapefinders. 
Adding statistical tools to their machinery, they looked at the maximum length 
scale at which filaments in that survey are statistically significant. They found 
that scale to be 50\,$h^{-1}$\,Mpc to 70\,$h^{-1}$\,Mpc. Another very promising
approach has been suggested by \citet{stoica04}. They introduce 
another method to actually detect filaments. Their method is based on an
algorithm that has been successfully used to detect road networks, and 
they showed that it works very well for 2D mock data sets.

In this work, we will restrict our focus mainly to filaments. In slices through 
N--body simulations \citep[see for example][]{jenkins98}  
filaments appear to be very common. There is some variety in their sizes and
appearance. The most massive haloes are usually connected by very prominent 
filaments (see Figure~\ref{fig:slice}). However, some of the filaments could 
be sheets and only appear to be filamentary due to projection effects. Less dense 
and less massive filaments can be found in less dense regions -- those filaments 
resemble fine perl necklaces. In an earlier work \citep{colberg99}, we showed that 
the formation of clusters is intimately linked with the cosmic neighbourhood. 
Infall of matter into the region where a cluster is forming happens from a few 
preferred directions. These agree with the locations of filaments at $z=0$.

Here, we will investigate the properties of filaments in more detail. In particular, 
we will address the following set of questions: On average, how many filaments 
intersect at the location of a cluster? Does the number of filaments depend on 
the mass of the cluster? What is the typical density of a filament? How long are 
these filaments typically? For a pair of clusters at some separation what is the
likelihood of finding a filament between them?

This paper is organized as follows: After a brief discussion of earlier
theoretical work (\ref{earliertheory}), in Section \ref{observations},
we discuss the status of observations to locate filaments. Afterwards (Section 
\ref{Filaments}), we briefly describe the simulation that we use (Section 
\ref{Simulation}) and then the procedure (Section \ref{Procedure}) to look at 
cluster--cluster connections. We present a classification of those 
connections in Section \ref{Classification}. Section \ref{Properties} describes 
properties of filaments and of clusters connected to such filaments. In Section 
\ref{Properties} we also discuss what our results might mean for alignments 
of clusters. We conclude the paper with a summary (Section \ref{Summary}).

\section{Theoretical work} \label{earliertheory}

Anisotropic collapse of matter in gravitational instability scenarios has
been known to lead to the formation of sheets and filaments since the seminal
work by \cite{zeldovich70} \citep[also see][]{shandarin89}. \cite{icke73} 
looked at the effect using homogeneous ellipsoidal models \citep[also see 
the work by][]{white79,eisenstein95}. \cite{bond96} then emphasized the role 
of tidal fields and of anisotropic collapse in the formation of LSS. They 
connected tidal shear directly to the locations of filamentary structures and 
of peaks. 

That same year, \cite{vandeweygaert96} investigated the typical morphology of a configuration
of two clusters with material in between to find that the primordial shear
constraint naturally evolves into a configuration of two clusters that are
connected by a filament. \cite{vandeweygaert02} contains a very detailed
summary of these theoretical efforts. 

There have also been theoretical studies of gas in filaments, the so--called
warm--hot intergalactic medium (WHIM), using simulations. The WHIM contains
a significant fraction of all baryons in the present-day universe (about
30\% to 40\%), which makes observing filaments through the signature of their
baryons quite interesting. For the most recent studies see \cite{dave01},
\cite{kravtsov02}, and \cite{furlanetto03} and references therein.

\section{Observational status} \label{observations}

The questions we posed in the introduction are of particular interest given 
observational efforts to find filaments by looking at the space in the vicinities
of clusters. \citet{gray02} and \citet{dietrich04} looked at cluster pairs 
A901/A902 and A222/A223, respectively. Using weak lensing they both conclude that 
the two clusters in their respective surveys are connected by a filament. 
\citet{pimbblet04} used clusters A1079 and A1084 in a pilot study to look for 
filaments, reporting a ``filament detection at a 7.5$\sigma$ level.''  
\citet{tittley01} used x--ray data to detect a filament between clusters A3391 
and A3395. 

The most relevant detection of a filament that was not found by looking at 
clusters directly is by \cite{scharf00}. They found a ``5$\sigma$ significance 
half--degree filamentary structure'', present both in x--ray and optical data.
At a likely redshift of around $\gtrsim$\,$z=0.3$, \citet{scharf00} give the 
length of the structure as $\gtrsim$\,12\,$h_{50}^{-1}$\,Mpc. 

Superclusters are very likely locations of filaments or even sheets. For example, 
\citet{connolly96} found a large structure at a redshift of $z=0.54$ with an
overdensity (in galaxies) of about four that includes three X--ray clusters. The
galaxies in this sample form ``a linear structure passing from the Southwest of 
the survey field through to the Northeast.'' More recently, \citet{bregman04}
looked at filaments in superclusters through UV absorption line properties of three 
AGNs projected behind possible filaments in superclusters. They conclude that
their results are consistent with the presence of filaments.

At somewhat larger redshifts, \citet{gal04} investigated two clusters at redshifts 
of $a \approx\,0.9$ which appear to be connected by a large structure. 
\citet{ebeling04} detected a structure of galaxies extending out from the 
cluster MACS J0717.5+3745, that is located at a redshift of $z=0.55$, with a 
length of 4\,$h_{70}^{-1}$\,Mpc.

\cite{kaastra03} observed a sample of 14 clusters, looking for soft X--ray excess 
emission. For five of their clusters they find ``a significant soft excess,''
which they attribute to emission from intercluster filaments of the WHIM in the 
vicinity of these clusters.

All these works are very exciting and indicate that there will be many more
such projects in the very near future. We thus feel that answering the questions
about what one can expect to find is all the more relevant.

\begin{figure*}
\caption{A slice of thickness $10\,h^{-1}$\,Mpc through the GIF simulation. The dark 
         matter was smoothed adaptively, and the resulting density field is shown 
         using a logarithmic colour scale. We have marked the location of the filament
         that is shown in Figure~\ref{fig:warpedfilament} with a box. In order to
         emphazise the r\^ole of the clusters we have included the clusters inside
         the box. The width of the box -- its dimension perpendicular to the 
         cluster--cluster axis -- is $15\,h^{-1}$\,Mpc, which is the actual diameter
         of the cylinders used to cut out filaments. -- Download from 
         {\bf http://lahmu.phyast.pitt.edu/$\sim$colberg/GIF\_10\_42\_z.jpg}}
\label{fig:slice}
\end{figure*}

\begin{figure*}
  \begin{center}
    \begin{tabular}{cc}
      \begin{minipage}{111mm}
        \begin{center}
          \includegraphics[width=110mm]{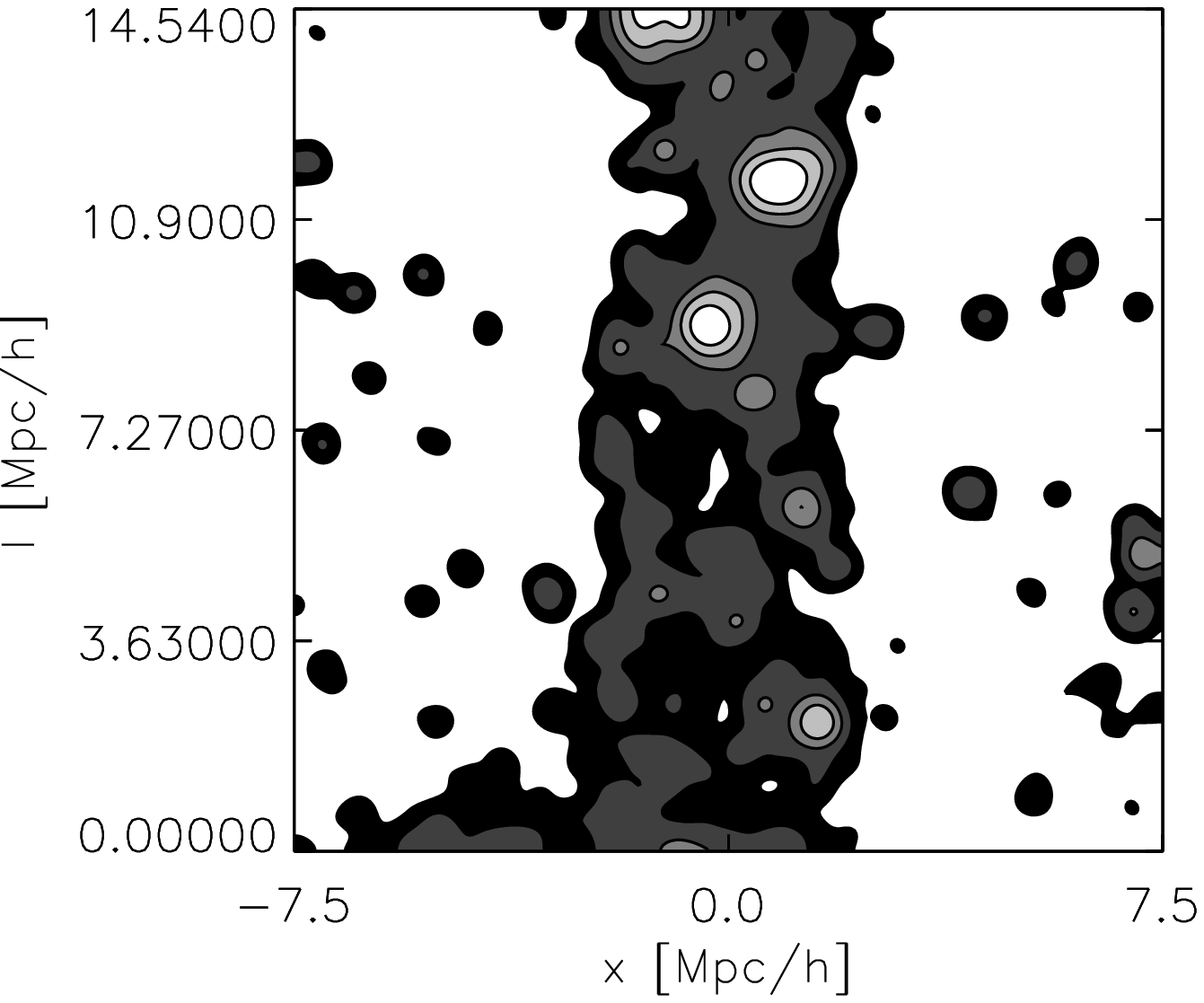}
        \end{center}
      \end{minipage}
      \hspace{-2.5cm}
      \begin{minipage}{111mm}
        \begin{center}
          \includegraphics[width=110mm]{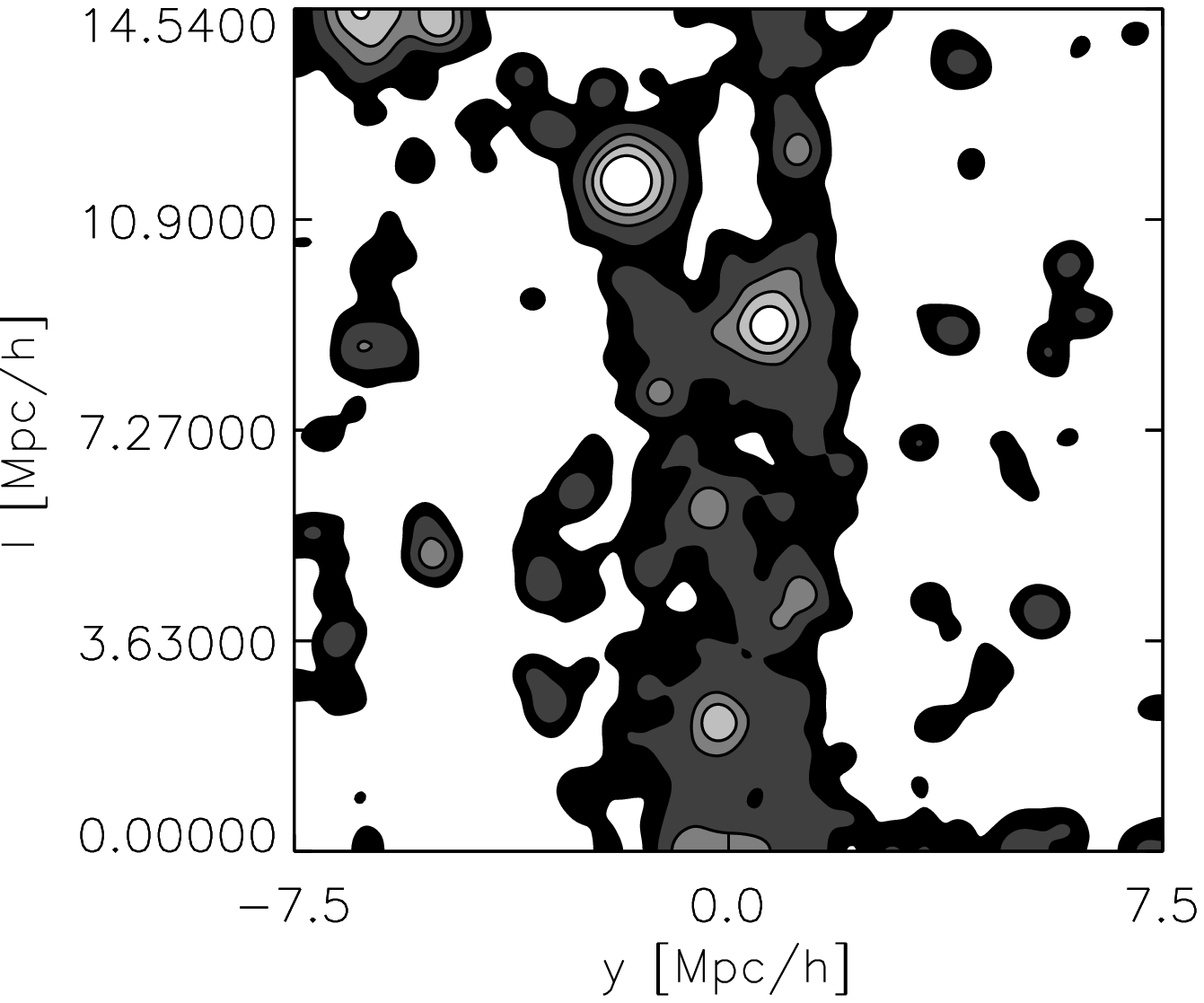}
        \end{center}
      \end{minipage}
    \end{tabular}
    %\vspace{-5mm}
  \end{center}
  \caption{Two orthogonal projections of the dark matter between two of
           the clusters in the GIF simulation. The plots show the projected
           overdensities, smoothed with a Gaussian of radius 0.5\,$h^{-1}$\,Mpc.
           The contour levels show overdensities ranging from 0.0 (mean density,
           black) to 19.0 (white). The $y$--axis cuts through both cluster
           centers. The region shown here excludes the clusters themselves.
           Matter follows a filamentary pattern.}
  \label{fig:filament}
\end{figure*}

\begin{figure*}
  \begin{center}
    \begin{tabular}{cc}
      \hspace{1.5cm}
      \begin{minipage}{111mm}
        \begin{center}
          \includegraphics[width=110mm]{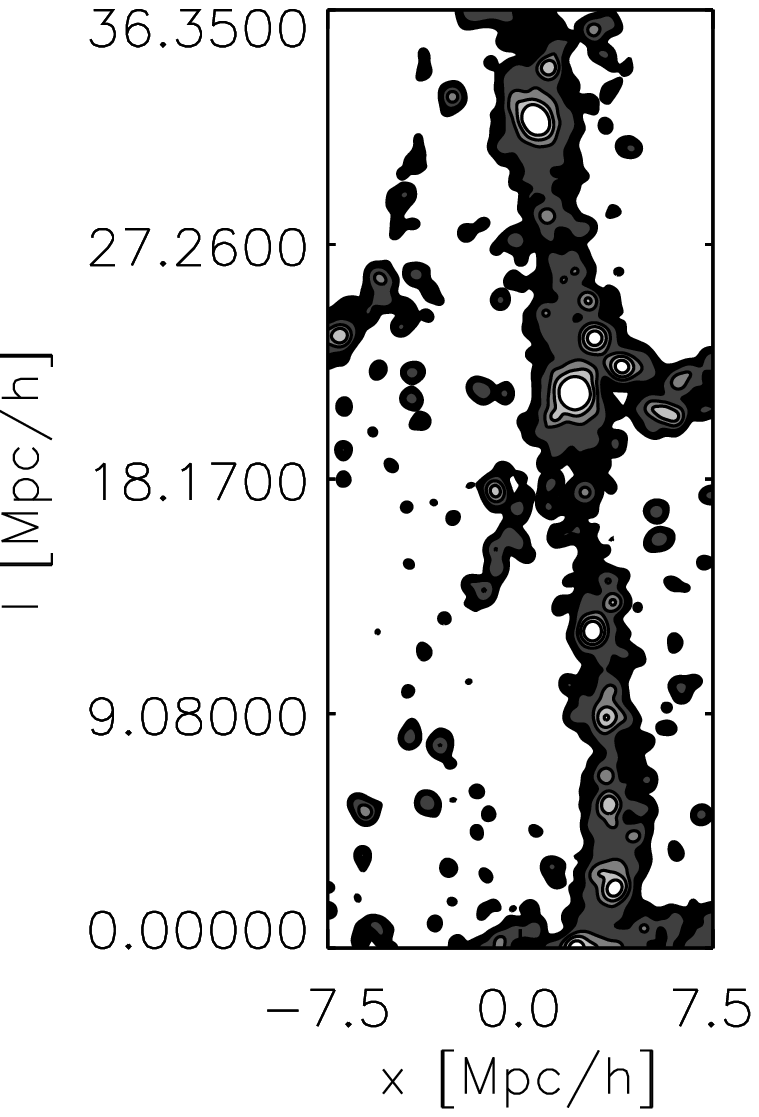}
        \end{center}
      \end{minipage}
      \hspace{-2.5cm}
      \begin{minipage}{111mm}
        \begin{center}
          \includegraphics[width=110mm]{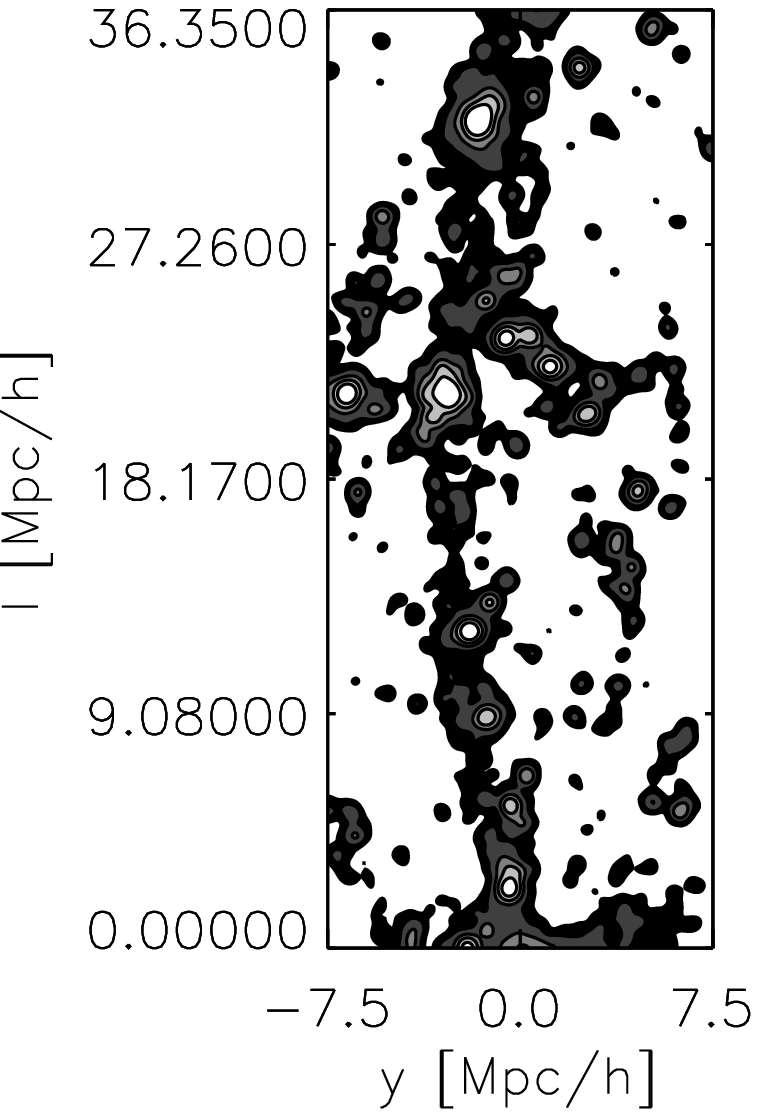}
        \end{center}
      \end{minipage}
    \end{tabular}
    %\vspace{-5mm}
  \end{center}
  \caption{Same as Figure \ref{fig:filament} for a moderately warped filament. Please note
           that this cluster--cluster connection is longer than the one shown in
           Figure \ref{fig:filament}.}
  \label{fig:warpedfilament}
\end{figure*}

\begin{figure*}
  \begin{center}
    \begin{tabular}{cc}
      \hspace{1.5cm}
      \begin{minipage}{111mm}
        \begin{center}
          \includegraphics[width=110mm]{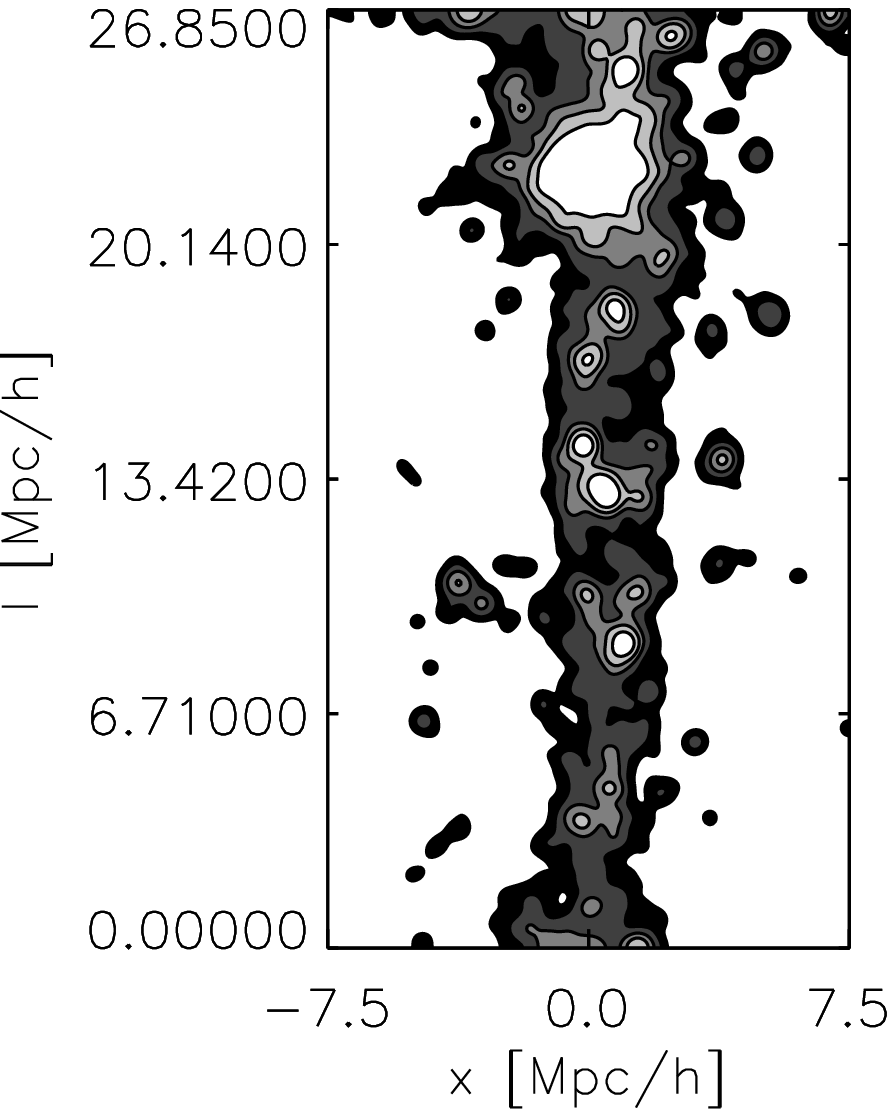}
        \end{center}
      \end{minipage}
      \hspace{-2.5cm}
      \begin{minipage}{111mm}
        \begin{center}
          \includegraphics[width=110mm]{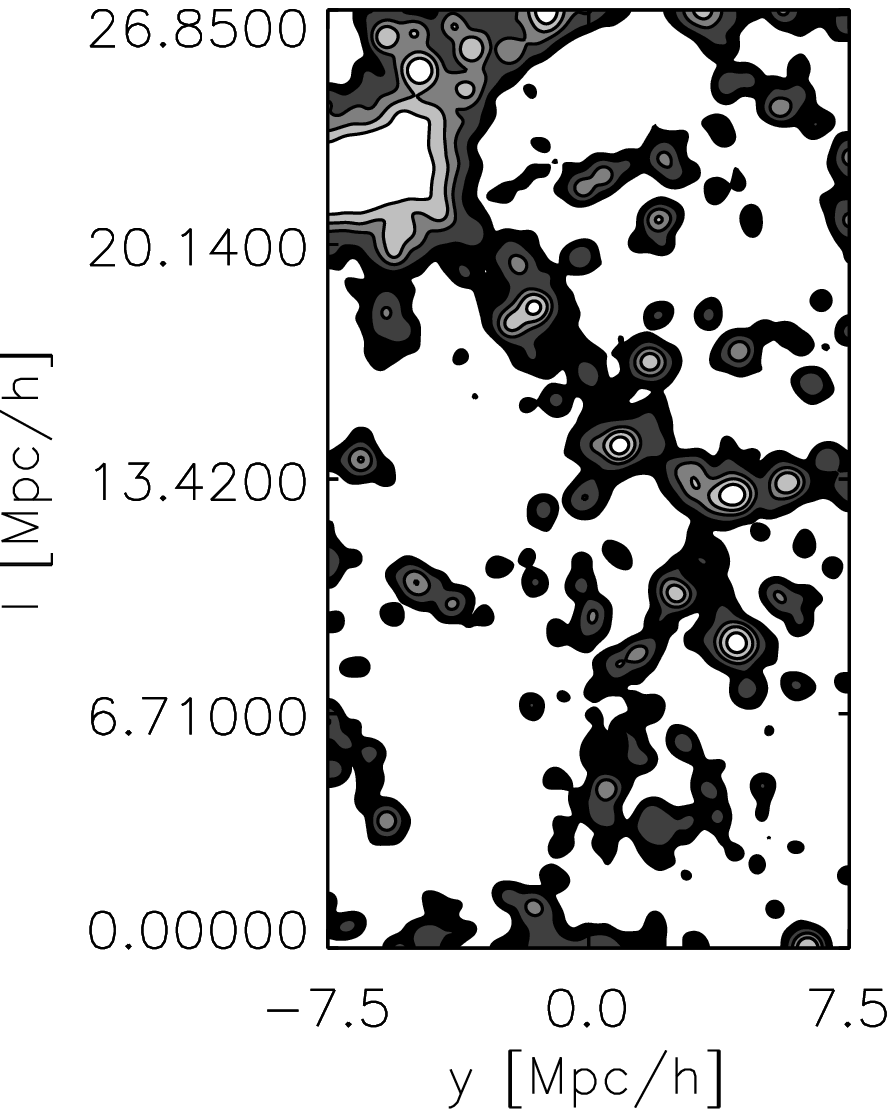}
        \end{center}
      \end{minipage}
    \end{tabular}
    %\vspace{-5mm}
  \end{center}
  \caption{Same as Figure \ref{fig:filament} -- again with a longer connection -- 
           for matter lying in a sheet. Note the presence of another cluster inside 
           the sheet but outside of the 5\,$h^{-1}$\,Mpc exclusion zone (see discussion).}
  \label{fig:sheet}
\end{figure*}

\begin{figure*}
  \begin{center}
    \begin{tabular}{cc}
%      \hspace{1.5cm}
      \begin{minipage}{111mm}
        \begin{center}
          \includegraphics[width=110mm]{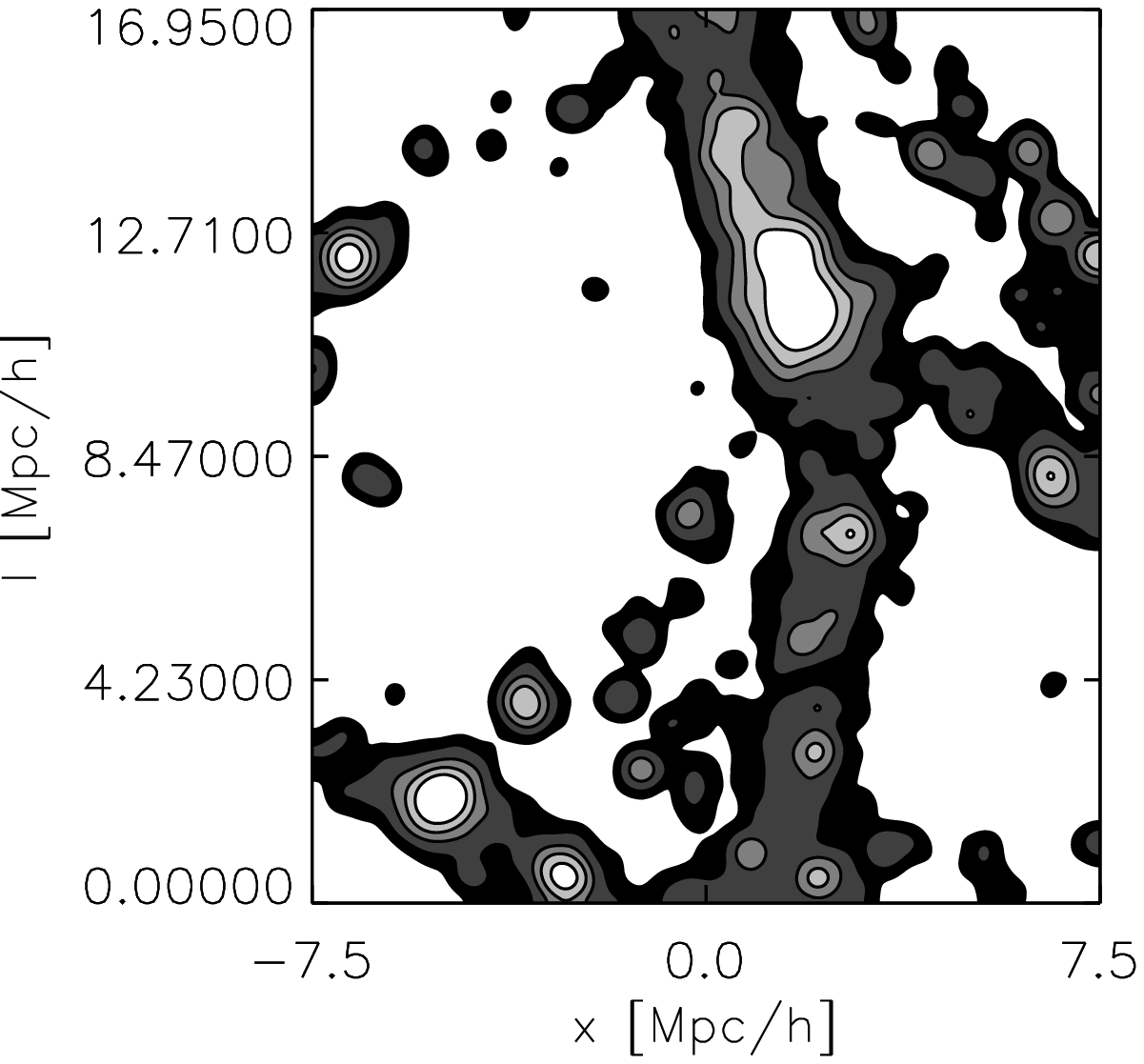}
        \end{center}
      \end{minipage}
      \hspace{-2.5cm}
      \begin{minipage}{111mm}
        \begin{center}
          \includegraphics[width=110mm]{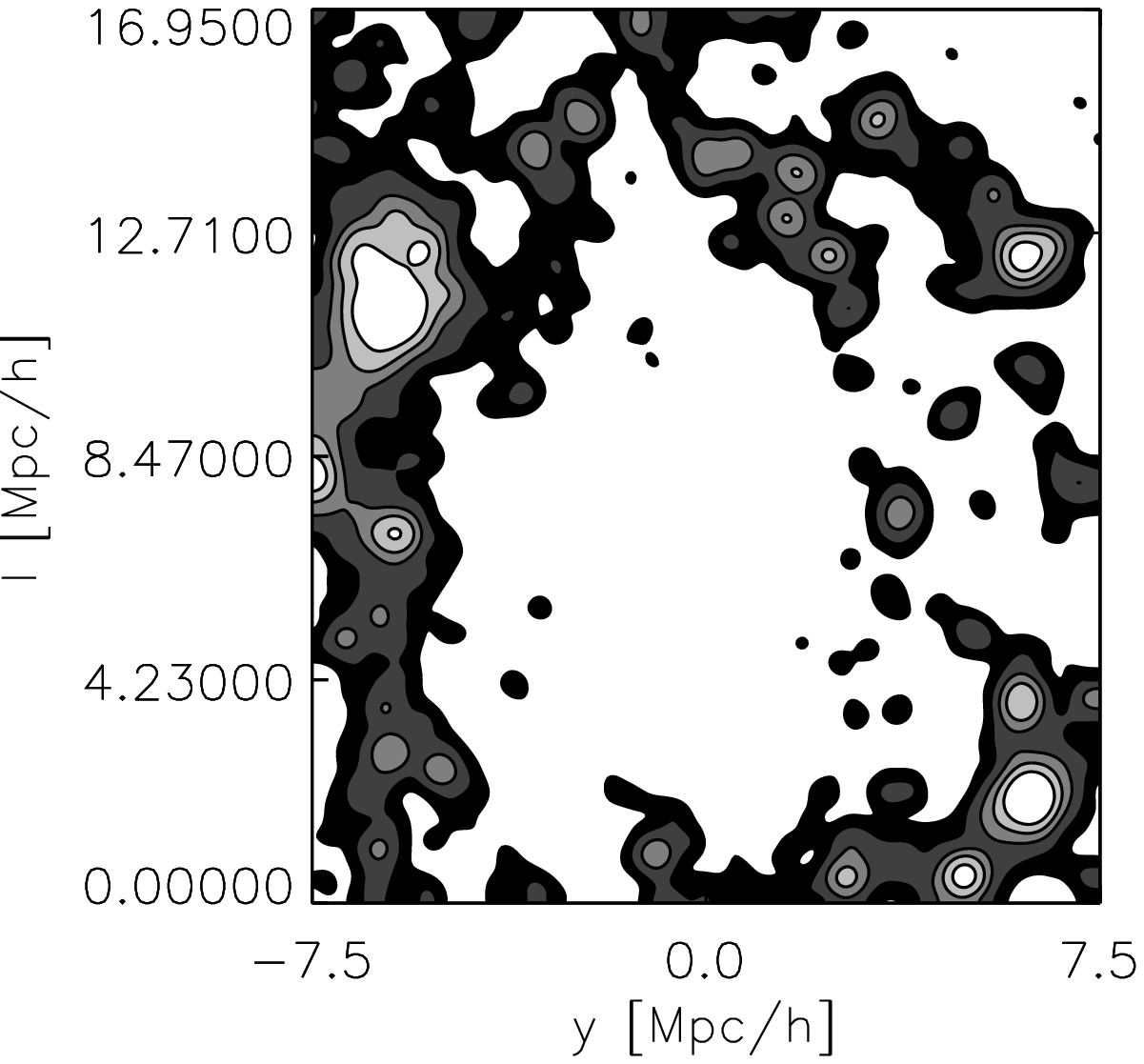}
        \end{center}
      \end{minipage}
    \end{tabular}
    %\vspace{-5mm}
  \end{center}
  \caption{Same as Figure \ref{fig:filament} -- again with different length scale -- 
           for matter that appears to be filamentary from one direction (left panel). 
           Viewed from another direction -- on the right--hand side from a direction 
           perpendicular to the one on the left--hand side there is no visible
           filament. The filament on the left--hand side is merely caused by a 
           projection effect.}
  \label{fig:projection}
\end{figure*}

\section{The inter--cluster network of matter} \label{Filaments}

\subsection{The simulation} \label{Simulation}

For this work, we use the $\Lambda$CDM simulation introduced in \citet{kauffmann99}. 
The simulation parameters correspond very closely to what has recently 
become the standard cosmology with 30\% of the critical density contributed by 
Cold Dark Matter and the remaining 70\% by a cosmological constant (or ``Dark Energy''). 
A Hubble constant of $h=0.7$\footnote{Throughout this work, we will express the Hubble 
constant in units of $H_0 = 100 h\,$km/sec/Mpc.} is used and the model is cluster 
normalized to $\sigma_8=0.9$. With $256^3$ particles in a box of size 
(141.3\,$h^{-1}$\,Mpc)$^3$ the simulation volume is large enough to study large--scale
structure at high resolution\footnote{The simulation data and halo catalogue can 
be downloaded from {\tt http://www.mpa-garching.mpg.de/galform/virgo/} {\tt hrs/index.shtml}.}.

\subsection{Visual impression} \label{visualimpression}

Figure~\ref{fig:slice} shows a slice of thickness $10\,h^{-1}$\,Mpc through the simulation
volume. The dark matter distribution was smoothed adaptively, and the resulting density field 
is shown using a logarithmic colour scale. The slice shows the network of matter, which is
quite familiar from simulation work and from galaxy redshift surveys. 

We have marked the location of the filament that is shown in Figure~\ref{fig:warpedfilament} 
with a box. In order to emphazise the r\^ole of the clusters we have included the clusters 
inside the box (Figure~\ref{fig:warpedfilament} shows just the material inside the filaments,
excluding the clusters). The width of the box -- its dimension perpendicular to the 
cluster--cluster axis -- is $15\,h^{-1}$\,Mpc, which is the actual diameter of the cylinders 
used to cut out filaments. Figure~\ref{fig:slice} gives a good impression of the extent of the 
longer filaments in our sample\footnote{Note that the large halo that is close to the cluster at
the top of the box has a mass below our cluster threshold mass. See Section \ref{Procedure} 
for details.}.

\subsection{The procedure} \label{Procedure}

First, we extract clusters from the halo catalogue that we downloaded from the
website of the Max--Planck--Institut f\"ur Astrophysik\footnote{{\tt 
http://www.mpa-garching.mpg.de/galform/virgo/index.shtml}}. We consider all
haloes more massive than 10$^{14}$\,M$_\odot$. This provides a sample of 170 
clusters whose number density is roughly comparable to that of R=0 Abell clusters 
\citep{postman92}.

Since we are interested in the inter--cluster network of matter, we examine  
the twelve nearest neighbours of each cluster. For each pair of clusters we extract the 
dark matter between them as follows. The cluster centers define an axis. We extract all 
dark matter particles in a cylinder of radius 7.5\,$h^{-1}$\,Mpc around
the axis. Furthermore, we only work with those particles that, when projected onto the axis, 
lie outside both clusters' $r_{200}$. That way, we do not extract  matter that belongs to either 
cluster. However, matter that lies in the cluster--infall regions is included in the samples. 
The value of 7.5\,$h^{-1}$\,Mpc is empirical (as is the number of twelve cluster neighbours). We 
found that going further out does not add any extra information whereas using smaller radii 
sometimes tended to cut warped filaments in half.

There is one final requirement for how neighbouring clusters were picked. Because we are 
interested in the inter--cluster network of matter we want to avoid finding one cluster 
lying directly between two other clusters. We thus do not allow another cluster to lie
inside the innermost 5\,$h^{-1}$\,Mpc from the cluster--cluster axis. As before, this value is 
empirical.

We want to emphasize that most certainly there will be many configurations where a number of 
clusters are lined up on a long filament. We exclude these cases for the simple and only 
reason that we want to study the distribution of matter between two neighbouring clusters. 
In a follow--up study, we will come back to looking at the alignment of clusters on larger scales.

\subsection{Classifying inter--cluster connections} \label{Classification}

Conventional wisdom says that if one picks two neighbouring clusters, they are  
connected by a filament, they lie in a sheet, or there is a void between them. In reality, 
the number of possibilities is slightly more complicated. The following list of the 
configuration of matter between neighbouring clusters is based on the visual inspection of 
each of the cluster--cluster connections. We tried to automate the process by using density 
measures but there are too many cases with deviations from the simple configurations mentioned 
above.

\begin{description}
  \item {\bf Filaments:} 19\% of the 1207 cluster--cluster connections\footnote{Usually, 
    if cluster $n_1$ has cluster $n_2$ in its list of neighbours, $n_1$ also shows up in 
    $n_2$'s neighbour list. However, this is not always the case. Thus, the list of 
    connections deviates from the completely symmectric case $170\times 12/2 = 1020$.} 
    contain a filament. We found three different possible configurations:
    \begin{description}
      \item {\bf straight:} 38\% of all filaments are straight and on center with
        respect to the cluster--cluster axis.That is, the clusters lie on the axis 
        of the filament. See Figure \ref{fig:filament}.
      \item {\bf off center:} Another 9\% of all filaments are also fairly straight 
        filaments but their central axes do not align with the axis that connects the 
        cluster centers.
      \item {\bf warped/irregular:} 53\% of all filaments are not straight but are either 
        warped or consist of multiple parts. Warped filaments sometimes indicate the presence
        of another cluster that lies just outside the 5\,$h^{-1}$\,Mpc exclusion zone. When we
        looked at the matter distribution of warped filaments going beyond the cylinder 
        used for the analysis we found many cases where nearby mass concentrations must
        have tidally interacted with the  filaments. See Figure \ref{fig:warpedfilament}.
    \end{description} 
  \item {\bf Sheets:} In about 2\% of all cases, we found a sheet--like configuration between 
    cluster pairs. Sheets display a narrow extent when viewed edge on and a broad 
    uniform distribution of matter otherwise. See Figure \ref{fig:sheet}. 
  \item {\bf The Rest:} The remaining 79\% of cluster--cluster pairs do not fall into the 
    aforementioned categories. However, there are some interesting cases left. In 3\% of all 
    cluster pairs we found a large amount of matter between them that does not look either 
    filamentary or sheet--like. Instead, viewed from any angle matter fills the space between 
    the clusters almost uniformly. In 19\% of all cases that were not classified as
    a filament, sheet or crowded field projection effects lead to the appearance of filaments.
    When viewed from one angle there seems to be a filament, whereas viewed from another angle there 
    is none (see Figure \ref{fig:projection}). There is the possibility of a sheet being
    mis--classified as a projection effect or vice versa -- especially since the classification
    is done by eye. However, since we focus our work on studying filaments this uncertainty 
    does not affect the bulk of our results. Projection effects and, to a lesser extent, sheets 
    have to be kept in mind when looking at 
    actual measurements of the distribution of matter between clusters by means of gravitational 
    lensing as results from these very different configurations will appear the same. 
\end{description} 

\subsection{Properties of inter--cluster filaments} \label{Properties}

Our sample of straight and warped filaments is large enough to allow more detailed studies of 
their properties. It is important to bear in mind the restrictions imposed on the analysis by how 
we selected the filaments. By filament we mean a filament that connects two clusters. The
filament could be a segment of a larger filament that connects more than two clusters. We do 
not attempt to study large filaments in more detail here. Whenever we are talking about filaments 
in the following we mean filaments between pairs of clusters.

\begin{figure}
\includegraphics[width=85mm]{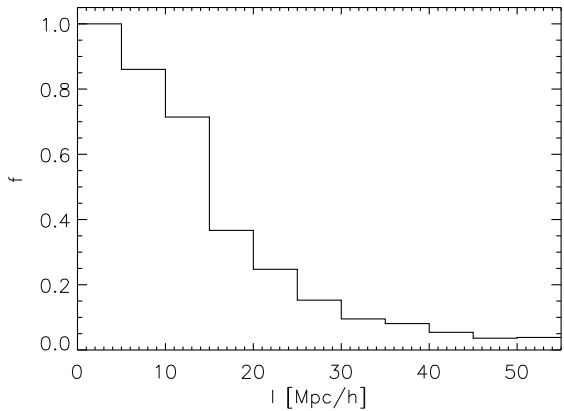}
\caption{Fractional abundance of filaments: The plot shows the fractions of cluster--cluster 
         connections that contain a filament -- straight or warped -- as a function of length of 
         those connections. Close cluster pairs are always connected by a filament.}
\label{FilamentLengthFraction}
\end{figure}

\begin{figure}
\includegraphics[width=85mm]{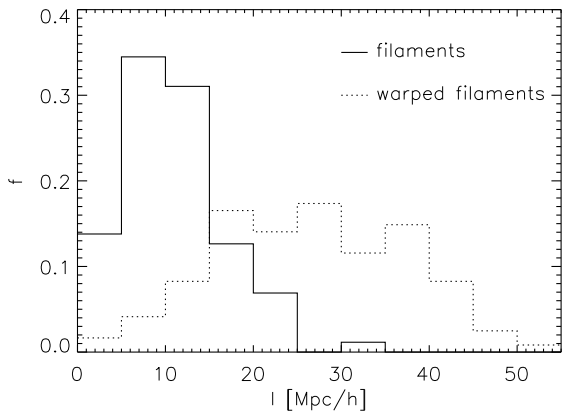}
\caption{Length distributions of filaments: The plot shows the fraction of filaments (solid line) 
         and warped filaments (dotted line) that have a length $l$. Warped filaments tend to be longer
         than straight filaments.}
\label{FilamentLengthDistribution}
\end{figure}

\begin{figure}
\includegraphics[width=85mm]{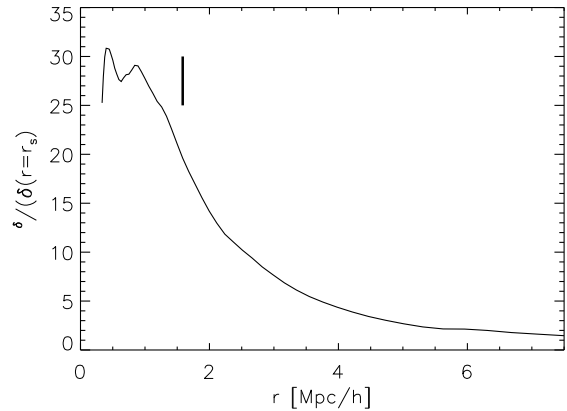}
\caption{Enclosed overdensity profile of a straight filament. The bold vertical line shows the radius  
         at which the profile starts to follow an $r^{-2}$ power law.} 
\label{DensityProfile}
\end{figure}

\begin{figure}
\includegraphics[width=85mm]{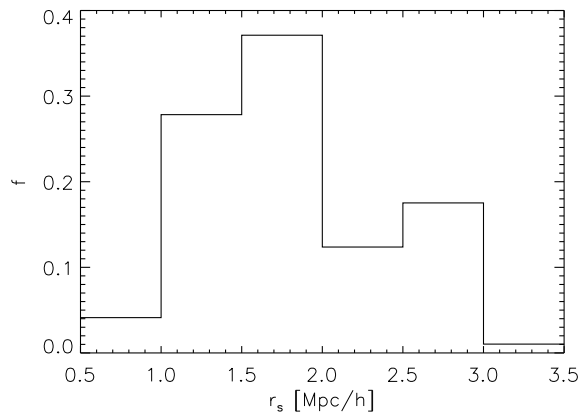}
\caption{Distribution of scale radii $r_s$ of straight filaments.}
\label{ScaleRadii}
\end{figure}

\begin{figure}
\includegraphics[width=85mm]{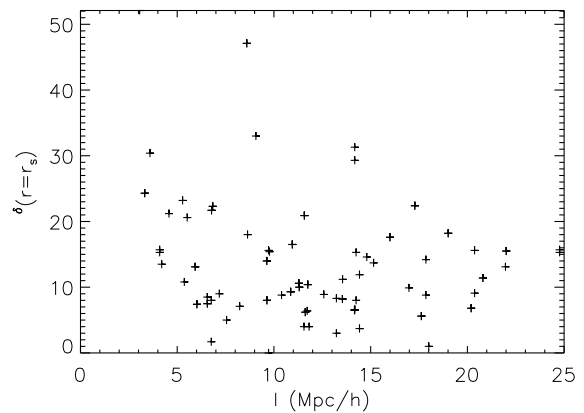}
\caption{For straight filaments, this plot shows the enclosed overdensity at scale radius $r_s$ 
         versus the length of the filament.} 
\label{delta_vs_length}
\end{figure}

\begin{figure}
\includegraphics[width=85mm]{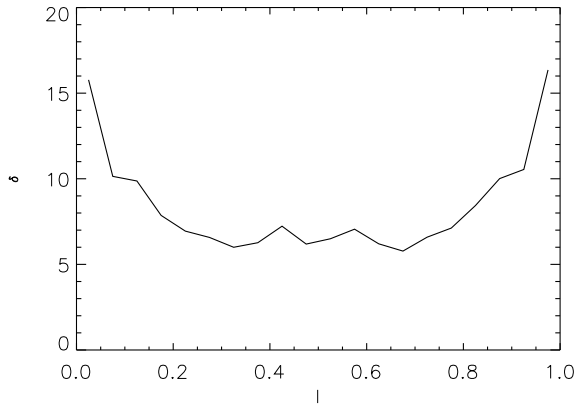}
\caption{Longitudinal density profile of straight filaments, averaged over straight filaments 
         that are longer than 5\,$h^{-1}$\,Mpc.
         Shown is the enclosed overdensity as a function of the positions along the cluster--cluster 
         axis for all material that is contained within 2\,$h^{-1}$\,Mpc from the axis.}
\label{DensityProfileLong}
\end{figure}

\subsubsection{General properties of cluster--cluster filaments} \label{PropertiesFilaments}

The first questions to ask about filaments are how long and how common they are. As we noted 
before, we looked at 1207 regions between neighbouring clusters of which 19\% contained a
filament. We selected the pairs of clusters by picking the twelve nearest neighbours of a 
given cluster. Given the fact that in the large--scale structure matter surrounds large voids, 
the relatively small number of filaments does not come as a surprise. If we had required that 
the cluster--cluster connections did not intersect voids we would have ended up with a larger 
overall percentage of filaments.

Nevertheless, it is interesting to look at the abundance of filaments for a given separation 
between two clusters. Figure \ref{FilamentLengthFraction} shows the fractions of cluster--cluster 
connections that contain a filament -- straight or warped -- as a function of the length of those 
connections. Very close cluster pairs are always connected by a filament, and about a third of 
all cluster pairs whose separation is between 15\,$h^{-1}$\,Mpc and 20\,$h^{-1}$\,Mpc are connected by a 
filament.

The fact that very close pairs of clusters -- with separations of up to 5\,$h^{-1}$\,Mpc -- are always 
connected by a filament can be explained by the presence of cluster infall regions. As discussed in, e.g.,  
Diaferio \& Geller 1997, the infall region of a cluster extends out to about three times its virial radius.
What this means is that two very close clusters do not constitute two strictly separated systems as 
their infall regions overlap, and if they are gravitationally bound, the two clusters may eventually
merge. Indeed, close cluster pairs show the presence of a filament between them \citep{dietrich04}.

Figure \ref{FilamentLengthDistribution} shows the length fractions of straight and warped 
filaments (strictly speaking, for warped filaments the length $l$ is not the length of the
filament but the length of the cluster--cluster connection). On the average, straight 
filaments are shorter than warped filaments. Two thirds of all straight filaments are in 
the 5\,$h^{-1}$\,Mpc to 15\,$h^{-1}$\,Mpc range whereas warped 
filaments have a much broader distribution that extends out to fairly large separations. 
Clearly, tidal fields must play a role here. In fact the visual impression is that many warped filaments 
are bent towards another mass concentration.

The most interesting properties of filaments are the amount of matter and its spatial
distribution inside the filament. We investigated these quantities for our sample of
straight filaments as follows\footnote{Unfortunately, warped and irregular filaments
cannot be investigated in such a straightforward fashion.}.  As noted above, all filaments
were identified visually. In almost all cases the axes of the straight filaments deviate
somewhat from the cluster--cluster axis. Therefore, for each filament we compute the actual axis 
by projecting all particles onto the plane perpendicular to the cluster--cluster
axis and then computing the center of mass of the particles in that plane. The
resulting center of mass is located where the filament's axis intersects with the
plane. We then use this axis to compute the enclosed (radial) mass profile by averaging
over all angles. 

The resulting profiles show a very interesting behaviour. There is a fairly large variation 
in the profiles close to the filaments' axes. However, for each filament we find that at 
some radius, $r_s$, the profile starts following an $r^{-2}$ power law closely. For 
each filament, we determine that radius, $r_s$, by finding the part of the profile that 
can be fit by an $r^{-2}$ power law. As an example, Figure \ref{DensityProfile} shows the 
radial overdensity profile of one filament. The bold vertical line marks $r_s$. 
The existence of the radius $r_s$ indicates that filaments have a well--defined edge.
With the mass of a filament $M_{fil}(>r_s) \approx const.$ but the volume growing as $r^2$ 
(the length $l = const.$) the enclosed overdensity of the filament $\delta(<r) \propto r^{-2}$. 

In Figure \ref{ScaleRadii} we plot the distribution of the radii $r_s$.
The majority of filaments posess radii $r_s$ between 1.0\,$h^{-1}$\,Mpc and 
2.0\,$h^{-1}$\,Mpc but there are also narrower and wider filaments. 

Given that filaments have well--defined edges how much matter is contained in a filament
and does the amount of matter depend on the length of the filament? Figure \ref{delta_vs_length} 
shows for each filament the enclosed overdensity at $r_s$ versus the length of the
filament. There is no obvious trend with length, but there is a fairly large scatter
in the enclosed matter. There are even a few cases where the enclosed overdensity
exceeds 30. We visually inspected these cases and found that for every such case there
is at least one very large halo present that lies on the filament's axis and whose mass
is below the cluster threshold mass used for this work. 

The radial density profiles of filaments have quite interesting repercussions for
observational efforts to find filaments. Given that filaments have well--defined
edges with no dependences on length and with fairly large possible masses finding
them observationally could be easier than previously thought. Either surveys using
gravitational lensing or direct observations of galaxies in the vicinity of galaxy
clusters \citep{ebeling04} appear to be very promising in this light.

Figure \ref{DensityProfileLong} shows the averaged longitudinal overdensity profile of
straight filaments. For these, we normalized the length of all filaments to unity and excluded
cluster pairs with separations less than 5\,$h^{-1}$\,Mpc. We then computed the enclosed density 
of all material that is contained within 2\,$h^{-1}$\,Mpc from the filament's axis as a 
function of the position along the axis. The overdensity rises towards the 
clusters and is constant at a value of around 7 in between the clusters. However, one needs 
to be careful with the profile for the following two reasons. First, individual
filaments show quite a bit of irregular lumpiness when plotted this way. Averaging over
the whole sample removes the lumpiness. Thus, while the averaged longitudinal profile is
fairly smooth, individual examples look vastly different. And second, the procedure we
use neglects the fact that infall regions have different sizes. We also plotted the 
averaged longitudinal profiles of subsamples of filaments whose lengths are comparable.
Because of the individual lumpiness of the filaments and because of the modest samples 
sizes the different sizes of the infall regions are lost because of the scatter of the 
data. Therefore, while averaging over all filaments longer than 5\,$h^{-1}$\,Mpc is not 
ideal, our sample does not allow us to do more detailed studies. 

\begin{figure}
\includegraphics[width=85mm]{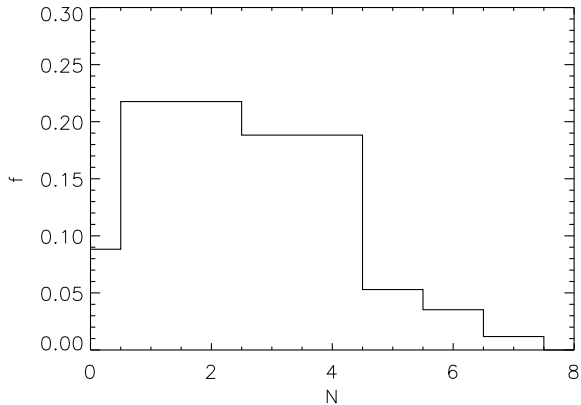}
\caption{Fractions of filaments for clusters: The plot shows the fractions of clusters that 
         have 0, 1, 2, ... filaments (both straight and warped). Less than 10\% of all 
         clusters are not connected to a filament.} 
\label{FilamentPerClusterFraction}
\end{figure}

\begin{figure}
\includegraphics[width=85mm]{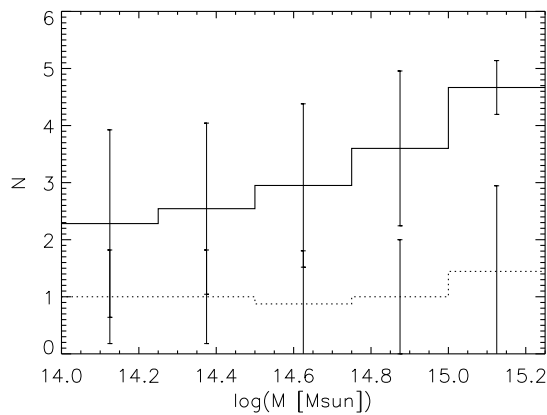}
\caption{Average number of filaments for clusters as a function of cluster mass. 
         Shown are clusters from the simulation (solid line) and clusters whose positions have been 
         randomized (dotted line -- see main text for details). Overplotted error bars show the 
         standard deviation. The number of filaments clearly increases as clusters get more massive.} 
\label{FilamentPerClusterMassFraction}
\end{figure}

\begin{figure}
\includegraphics[width=85mm]{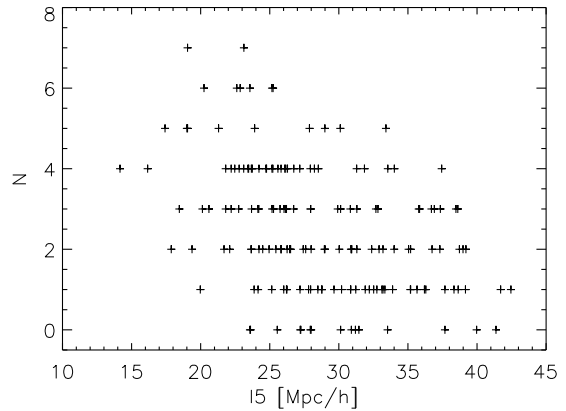}
\caption{For each cluster, this plot shows its number of filaments as a function of the distance 
         to its fifth--nearest neighbouring cluster.}
\label{NeighboursVsFilaments}
\end{figure}

\subsubsection{Clusters and filaments} \label{ClustersAndFilaments}

Since we started out with clusters to find filaments, it is quite natural to come back to them. 
The first obvious question to ask is how many filaments we can find per cluster. Figure 
\ref{FilamentPerClusterFraction} shows the distribution of filaments per cluster. The vast 
majority of clusters are connected to between one and four filaments. 

The average number of filaments per cluster is a somewhat unsatisfactory quantity on its 
own. It gets more interesting when one looks at the mean number of filaments per cluster as 
a function of cluster mass -- shown as the solid histogramme in Figure \ref{FilamentPerClusterMassFraction}. 
There is a clear trend with mass. More massive clusters are connected to more filaments. For 
the most massive clusters in the sample, the mean number of filaments is almost five. This 
picture agrees with the visual impression from simulations (see the images in \citet{jenkins98}  
and \citet{kauffmann99}) where the backbone of large--scale structure is formed by the most massive 
objects.

We tried to evaluate this result by randomizing the cluster positions and then looking
for filaments again. The cluster positions were changed as follows. We first produced a
smoothed version of the density field in the simulation volume, using a Top Hat with a
radius of 2\,$h^{-1}$\,Mpc. We then assigned the clusters randomly to cells whose overdensity
was five or more. This way, we made sure that the randomized clusters ended up somewhere
in the overdense parts of the simulation volume and not in a void. Having obtained this
new set of clusters we re--did the investigation of the cluster--cluster connections using
the same procedure as for the original cluster sample. The dotted histogramme in 
Figure \ref{FilamentPerClusterMassFraction} shows the result. As can be seen, the 
randomized clusters deviate from the original ones. The average numbers of filaments
per cluster are lower than those of the original clusters and there is no trend with
mass.

Using a scatter plot, Figure \ref{NeighboursVsFilaments} shows the number of filaments per 
cluster versus the distance to the clusters' fifth--nearest neighbour. The latter is
commonly used as a somewhat crude measure of density. There is a trend for clusters with 
closer neighbours to have more filaments. This can be understood from the preceding: We have 
already seen that more massive clusters connect to more filaments. And we also know that 
in CDM universes more massive objects are more strongly clustered than less massive objects. 

\begin{figure}
\includegraphics[width=85mm]{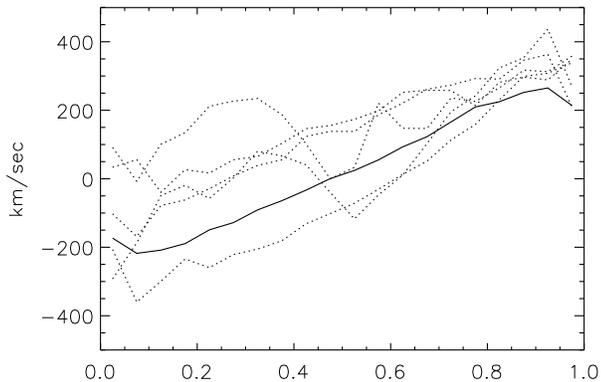}
\caption{Longitudinal velocity profile of straight filaments, averaged over straight
filaments that are longer than 5\,$h^{-1}$\,Mpc (solid line) and five longitudinal 
velocity profiles of individual filaments (dotted lines). Shown is the enclosed
mean velocity as a function of the positions along the cluster--cluster axis for all
material that is contained with 2\,$h^{-1}$\,Mpc from the axis. The velocities are
negative for material that moves towards the cluster on the lefthand side and positive
for material that moves towards the cluster on the righthand side.}
\label{vel_z}
\end{figure}

\begin{figure}
\includegraphics[width=85mm]{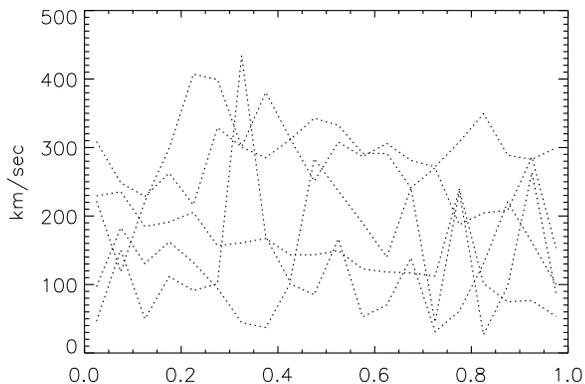}
\caption{Transversal velocity profiles of the five straight filaments, which were shown
in Figure \ref{vel_z} (dotted lines). Shown is the enclosed mean velocity as a function 
of the positions along the cluster--cluster axis for all material that is contained 
with 2\,$h^{-1}$\,Mpc from the axis.}
\label{vel_xy}
\end{figure}

\subsubsection{Velocity fields} \label{VelocityFields}

As a consequence of the role of filaments in the context of the formation of galaxy
clusters (see \citet{colberg99}) it is most natural to assume that the velocity field
at any given point inside a filament is determined by the distance to the two clusters.
In particular, material probably moves towards the cluster it is nearest to. We use the
straight filaments in our sample to investigate this. In order to avoid contamination
through very close clusters pairs, which might be merging or whose infall regions
overlap, we only consider cluster pairs that are separated by 5\,$h^{-1}$\,Mpc plus
their respective virial radii.

Figure \ref{vel_z} shows the averaged longitudinal velocity profile of straight filaments (solid
line) and the longitudinal velocity profiles of five individual straight filaments (dotted lines). 
For the profiles, we normalized the length of all filaments to unity. We then computed the enclosed 
mean velocity of all material that is contained within 2\,$h^{-1}$\,Mpc from the filament's 
axis as a function of the position along the axis. The velocities are negative for material 
that moves towards the cluster on the lefthand side and positive for material that moves 
towards the cluster on the righthand side. As can be seen, the average profile very clearly 
shows the gravitational domains of the two clusters. On the average, material tends to move 
towards the nearest cluster. The individual profiles, however, show a fair amount of scatter. 
While the overall trend is the same as for the average profile, individual longitudinal velocity 
profiles are not nearly as smooth as the average. This fact can be understood from the clumpy 
structure of filaments. As seen above, the material in filaments is not distributed smoothly. 
Instead, single haloes determine the structure of a filament, with fairly large differences 
between individual filaments. Figure \ref{vel_xy} shows this very clearly. Here, we plot the 
transversal velocity profiles of the five straight filaments, which were shown in Figure 
\ref{vel_z}. The scatter between the filaments is very large, and it is easy to make out individual 
haloes.

\citet{eisenstein97} proposed a method to measure the mass of filaments by studying their transverse 
velocity dispersion. We intend to investigate this
method in an upcoming paper where we make use of a much larger simulation, which, with about 600 
times as many particles in a volume about 44 the size of the simulation used here, has many more 
filaments and a much better mass resolution.

\begin{figure}
\includegraphics[width=85mm]{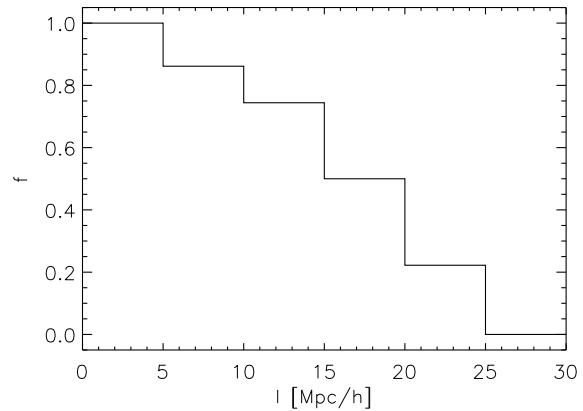}
\caption{Fractions of clusters for which there is a filament in the connection to the nearest neighbour
         as a function of the distance.}
\label{Bingelli}
\end{figure}

\subsubsection{Cluster--cluster alignments} \label{ClusterClusterAlignments}

For a long time, there has been discussion on whether clusters of galaxies are
aligned with their neighbours (see \citet{binggeli82} for the original work and, e.g.,
\citet{chambers00} and \citet{plionis02} for recent updates). 
\citet{onuora00} studied alignments of clusters in large simulations and 
found ``strongly significant alignments'' for separations of up to 30\,$h^{-1}$\,Mpc 
in the $\Lambda$CDM model \citep[also see][]{faltenbacher02}. As an 
explanation of alignment -- if the effect exists -- filaments have been brought 
up. \citet{west95} suggested cluster formation along filaments 
\citep[compare][]{colberg99} had implications on the orientations of clusters. 
There are two possible explanations for this effect. First, the primordial
density field pre--determines the directions from which matter falls into
clusters. There is a positive correlation between the inertia tensor of a
cluster and its surrounding tidal field \citep[see][]{bond96}. Second, as
shown in \cite{vanhaarlem93}, clusters tend to orient themselves towards
the direction of last matter infall. This finding, combined with the results
of \cite{colberg99} suggests that we can expect neighbouring clusters to be aligned 
{\it if} the formation of clusters along filaments is the dominant factor, which 
decides about alignment, and {\it if} there actually is a filament between the clusters. 

We have already seen earlier (Section \ref{ClustersAndFilaments}) that not all 
neighbouring clusters are connected by a filament. Because alignment studies often 
focus on the nearest neighbour we want to shed additional light on this point in 
this context. Figure \ref{Bingelli} shows the fractions of clusters for which there 
is a filament in the connection to the nearest neighbour plotted against the length 
of the connection. Note that this is somewhat of a variation of Figure 
\ref{FilamentLengthFraction}. Instead of looking at all cluster--cluster connections, 
we look only at those for the nearest neighbour of each cluster. As before, 
we see that very close pairs of clusters are always connected by a filament. For 
larger separations, the fraction drops, but even for separations between 10\,$h^{-1}$\,Mpc 
and 15\,$h^{-1}$\,Mpc, the likelihood of finding a filament towards the nearest neighbour 
of a cluster is around 75\%. Therefore, if the alignment of neighbouring clusters is caused 
by matter infall from filaments, there is a very high chance of finding cluster 
alignment up to separations of 15\,$h^{-1}$\,Mpc. Going further out, the likelihood 
drops quite steeply. 

There is another consequence of what we just discussed. If the formation of clusters 
along filaments is responsible for alignment, the non--detection of alignment of 
neighbouring clusters -- separated by 5\,$h^{-1}$\,Mpc or more -- does not necessarily mean 
that there is no such effect. It could simply mean that there is no filament between 
the clusters.

\subsubsection{Clusters and sheets} \label{ClustersAndSheets}

As noted earlier, we were able to find a few sheet--like connections between clusters. We 
want to emphasize first that, as for the case of the filaments we discuss, we are only 
looking at sheet--like configurations of matter between neighbouring clusters. Whether or 
not some clusters lie in larger sheets -- in galaxy redshift surveys these sheets are 
usually called walls -- is beyond the scope of this work. We also note that 
some of the sheets could be mere projection effects and we might have missed some sheets 
and classified them as projection effects. Given this uncertainty and given the small
number of sheets we do not look at sheets in too much detail.

Sheets appear to be much rarer than filaments. We found a total of less than two dozen 
sheets. In almost all cases there was only one sheet per cluster. The mean length of 
the connections that contain a sheet is 27.4\,$h^{-1}$\,Mpc. We did not find very short 
sheets -- with lengths smaller than 10\,$h^{-1}$\,Mpc -- or very long sheets -- with lengths 
exceeding 45\,$h^{-1}$\,Mpc. 

Sheets appear to have lower surface densities than filaments. One might argue that
the mass resolution of the simulation is not good enough to properly address the
issue of how many sheets there really are. While we think that the mass resolution
of our simulation definitely is high enough and that sheets indeed are not as common
as filaments we want to note that we intend to re--address this issue in later
work that will make use of a larger simulation with much higher mass resolution.

\section{Summary} \label{Summary}

Using a high--resolution N--body simulation of cosmic structure formation in a
$\Lambda$CDM Universe, we studied the material between pairs of clusters that
correspond to R=0 Abell clusters. Our main results can be summarized as follows:
\begin{itemize}
  \item Whereas sheets appear to be fairly rare, filaments between clusters are
        very common. The likelihood of finding a filament between two neighbouring
        clusters increases as the separation of the clusters decreases. Very close
        pairs of clusters that are separated by 5\,$h^{-1}$\,Mpc are always connected
        by a filament. Warped or irregular filaments are more common than straight
        filaments -- due to the presence of tidal fields between clusters. The longer
        the cluster--cluster connection the higher the likelihood is of finding a
        warped or irregular filament rather than a stright one.
  \item We investigated straight filaments in more detail to look at the amount
        of matter in those filaments. Filaments are very lumpy objects. The longitudinal
        overdensity profile -- measured along the axis that connects the two clusters --
        clearly shows this. Towards the two clusters the overdensity rises as the
        cluster infall regions are reached. The radial enclosed overdensity profiles of
        filaments show a well--defined radius at which the profiles follows
        an $r^{-2}$ profile. For the majority of filaments, this radius is between 
        1.0\,$h^{-1}$\,Mpc and 2.0\,$h^{-1}$\,Mpc but there are also narrower and 
        wider filaments. The enclosed overdensity inside this radius varies between
        a few times up to 25 times mean density or more. All high--density cases
        could be visually identified as containing large haloes. 

        The results from the density profiles indicate that finding filaments
        observationally might be somewhat easier than previously thought especially
        if the line of sight is aligned with the axis of a filament. Finding
        filaments that are perpendicular to the line of sight is trickier because
        of the lumpiness of filaments. It is probably most promising to look at
        the immediate vicinities of clusters where filaments have higher densities.
  \item The majority of all clusters possess between one and four filaments. There
        is a very clear trend with mass. More massive clusters on average have
        more filaments. This result supports the general view that the most massive
        clusters sit at the intersections of the backbone of large--scale structure.
        There also is a weak trend for clusters in denser regions to have more filaments.
        The last statement is not completely independet of the former one. More massive 
        clusters can be found in denser regions and are also more clustered.
  \item The velocity field in a filament is dominated by the two clusters. While there
        is a large scatter between individual filaments, material tends to move
        towards the clusters that is closest.
  \item Filaments have been used to explain alignments between neighbouring clusters.
        We find that there is a very high likelihood of finding a filament in the
        connection to nearest neighbour of a cluster for separations of up to 
        15\,$h^{-1}$\,Mpc. If filaments are indeed responsible for alignments one would
        expect to find closer pairs of clusters to be aligned. However, even at
        separations of between 5 and 15\,$h^{-1}$\,Mpc there are pairs of clusters
        that are not connected by a filament. Close clusters that are not alligned 
        thus could indicate that there is no filament between them.
  \item We find that the fraction of matter configurations that appear to be filamentary
        due to projection effects is about the same as the fraction of genuine
        filaments. This effect has to be taken into account when investigating the
        distribution of matter between neighbouring clusters observationally. The
        contribution of projection effects become non--negligible at separations of
        10\,$h^{-1}$\,Mpc.
\end{itemize}

The results of this paper have found application in \citet{pimbblet04b}, who
analyze the frequency and distribution of intercluster galaxy filaments selected from 
the 2dF Galaxy Redshift Survey.

\section*{Acknowledgments}

JMC and KSK acknowledge partial support through research grants ITR AST0312498
and ITR ACI0121671; AJC acknowledges partial support through research grant
NSF CAREER AST9984924.

Part of this research was triggered by talks and results presented at the
IAU Colloquium 195 in Torino, Italy, in March 2004. JMC thanks 
Antonaldo Diaferio for stimulating discussions during the workshop. We also 
thank the referee, Rien van de Weygaert, for his very helpful report and
Andrew Hopkins, Chris Miller and Ravi Sheth for their input and encouragement.

The simulation discussed here were carried out as part of the Virgo Consortium 
programme, on the Cray T3D/Es at the Rechenzentrum of the Max--Planck-Gesellschaft in 
Garching, Germany and at the Edinburgh Parallel Computing Center. We are indebted to 
the Virgo Supercomputing Consortium for allowing us to use it for this work. 

\bibliography{fil}
\label{lastpage}

\end{document}